\documentstyle[prl, aps]{revtex}      
\begin{document}                

\narrowtext 
\twocolumn

\title{{Absence of residual quasiparticle conductivity in the underdoped 
cuprate YBa$_{2}$Cu$_{4}$O$_{8}$}}
\author{N.E. Hussey}
\address{Department of Physics, Loughborough University, 
Loughborough LE11 3TU, U.K.}
\author{S. Nakamae, K. Behnia}
\address{UPR 5 Ð CNRS, ESPCI, 10 Rue Vauquelin, 75005 Paris, France.}
\author{H. Takagi, C. Urano}
\address{Department of Advanced Materials Science, Graduate School of Frontier 
Sciences, University of Tokyo, Hongo 7-3-1, Tokyo, 113-8656, Japan.}
\author{S. Adachi and S. Tajima}
\address{Superconducting Research Laboratory, International 
Superconductivity Technology Center, Shinonome 1-10-13, Tokyo 135, Japan.}
  
\maketitle
\begin{abstract}                
We report here measurements of the in-plane thermal conductivity 
$\kappa$ of the stoichiometric underdoped cuprate YBa$_{2}$Cu$_{4}$O$_{8}$ 
(Y124) below 1K. $\kappa$($\it T$) is shown to follow a simple, phononic $T^{3}$ 
dependence at the lowest $\it T$ for both current directions, with a 
negligible linear, quasiparticle contribution. This observation is 
in marked contrast with behavior reported in optimally doped cuprates, 
and implies that extended zero-energy (or low-energy) quasiparticles 
are absent in Y124.
\end{abstract}


Experimental evidence for $\it d$-wave superconductivity in 
high-$T_{c}$ cuprates is now well established. The presence 
of nodes in the gap is expected to produce a finite density of well-defined 
quasiparticle (QP) excitations at low energies that dominate
the low-$\it T$ physics. For a $\it pure$ $\it d$-wave superconductor 
with line nodes on the Fermi surface, for example, the density of states 
(DOS) is linear in the excitation energy, giving rise to a $T^{2}$ dependence 
of the low-$\it T$ specific heat and thermal conductivity (assuming a 
constant scattering rate). This excitation spectrum, however, is 
altered significantly in the presence of impurities. For an anisotropic 
two-dimensional (2D) superconductor with scattering in the unitary limit, a 
band of impurity states is expected to develop whose width $\gamma$ grows with 
increasing impurity concentration $n_{imp}$, leading to a finite zero-energy 
DOS\cite{gorkov,lee,hirsch,graf}. Lee showed that the residual conductivity 
is independent of $n_{imp}$, the result of compensation between the increased 
DOS and a reduction in the associated transport lifetime\cite{lee}. This 
residual, or "universal" conductivity develops at low $\it T$ in the 
so-called "dirty" limit, $k_{B}T \leq \gamma$. Similarly, the low-$\it T$ 
$\kappa(T)$ will be dominated by QP states in the vicinity of the line nodes, 
and is given by the expression\cite{graf}
\begin{equation}
    \kappa_{res}/T \approx (n/d) (k_{B}^{2}/ 3) (v_{F}/v_{2})	
\end{equation}
where $n/d$ is the stacking density of CuO$_{2}$ planes and $v_{F}$  and 
$v_{2}$ are the energy dispersions (QP velocities) perpendicular 
and tangential to the Fermi surface respectively.

The issue of localization of electronic states in $\it d$-wave 
superconductors, however, is a complex problem and many competing viewpoints  
prevail\cite{lee,hirsch,graf,balat,sent,vish,ners}. Balatsky and 
Salkola\cite{balat}, for example, argue that the long-range nature of hopping 
between impurity states along the nodal directions leads to strong overlap of 
the impurity wave functions along the diagonals of the square lattice, and 
ultimately to "extended" impurity states. Senthil and co-workers\cite{sent,vish}, 
on the other hand, argue that quantum interference effects destabilize the 
extended QP states and lead to a vanishing DOS at zero energy (See 
also\cite{ners} for a similar conclusion). They coined the term 
"superconducting insulator"\cite{vish} to describe such a superconductor 
with localized states.

Experimentally, the observation of a finite linear term in 
$\kappa(T)$ in pure and Zn-doped YBa$_{2}$Cu$_{3}$O$_{7-\delta}$ (Y123) by 
Taillefer $\it et$ $\it al.$\cite{taille} appeared to confirm the existence 
of zero-energy quasiparticles. Moreover, the size of this term was indeed 
found to be "universal", i.e. independent of Zn 
concentration, in agreement with Lee's prediction. Similar behavior was 
later reported for Bi$_{2}$Sr$_{2}$CaCu$_{2}$O$_{8}$ (Bi2212) by Behnia 
$\it et$ $\it al.$, using irradiated crystals\cite{behnia}. More recently, 
the magnitude of $\kappa_{res}/T$ in Bi2212 ($\approx$ 0.15 mW/cm.K$^{2}$) 
was shown to be consistent with absolute values of $v_{F}/v_{2}$ estimated 
from angle-resolved photoemission (ARPES)\cite{mesot,chiao}.

Despite this apparent consistency between theory and experiment, 
$\kappa_{res}/T$ has only been reported for two compounds, both at their 
optimum doping level, and it is not immediately obvious how $\kappa_{res}/T$ 
will vary across the phase diagram. ARPES\cite{mesot} and penetration depth 
measurements\cite{pana1} support claims that $v_{F}/v_{2}$, and thereby 
$\kappa_{res}/T$, increase in the underdoped regime. On the other hand, in 
certain underdoped cuprates, where $T_{c}$ has been suppressed in high 
magnetic fields\cite{ando,fournier}, there is a marked tendency towards 
localization below $T_{c}$, suggesting a $\it vanishing$ QP contribution at 
low $\it T$. Clearly, low-$\it T$ $\kappa(T)$ measurements on underdoped 
cuprates are important to help clarify this seemingly contradictory behavior.

With this in mind, we have carried out the first low-$\it T$ $\kappa (T)$ 
measurements on the underdoped cuprate Y124 ($T_{c}$ = 80K), which is a 
self-doped, stoichiometric cuprate and therefore 
relatively free of disorder. Below 0.25K, $\kappa (T) \approx T^{3}$ for 
both $\it a$- and $\it b$-axis currents, consistent with a phonon heat 
conduction in the ballistic regime. The residual linear QP term, however, 
is either absent or is negligibly small, with an upper bound of 0.02 
mW/cm.K$^{2}$. This result reveals that the universal conductivity scenario 
breaks down dramatically in underdoped Y124. One compelling possibility is that 
the QP states in Y124 are localized at low $\it T$, due 
to the proximity to the superconductor/insulator (S/I) boundary, and 
therefore do not contribute to the low-$\it T$ heat transport.

The Y124 crystals were grown by a flux method described 
elsewhere\cite{adachi}. For this particular study, three plate-like crystals 
were selected, two with their longest dimension along the $\it b$-axis, 
the other along the $\it a$-axis. Approximate dimensions were 0.25 x 0.16 x 
0.015 mm$^{3}$ for the $\kappa_{a}$ crystal (labelled hereafter as 
a$\sharp$1) and 1 x 0.09 x 0.05 mm$^{3}$ and 0.8 x 0.25 x 0.06 mm$^{3}$ for 
the two $\kappa_{b}$ crystals, b$\sharp$1 and b$\sharp$2. $T_{c}$ = 80K for 
all crystals,  with a transition width, measured resistively, of less than 1K.

$\kappa(T)$ for each crystal was measured between 0.14K and 1K using a 
conventional steady-state four-probe technique that allowed the electrical 
resistivity $\rho_{a,b}(T)$ of each sample to be measured in situ without changing 
the contact configuration. Gold wires were attached as electrical contacts using 
Dupont 6838 silver paint. The $\rho_{a,b}(T)$ behavior was found to be in excellent 
agreement with previous measurements\cite{huss}, with room temperature 
values, $\rho_{a}$ = 350 $\mu \Omega$cm and $\rho_{b}$ = 90 $\mu \Omega$cm 
(for both $\it b$-axis crystals). This large in-plane anisotropy arises 
from the high conductivity of carriers on the quasi-1D CuO chains 
that run parallel to the $\it b$-axis (see schematic inset to Figure 1) 
and confirms not only the high quality of the crystals used 
in this study, but also that current flow in each case is uniaxial. The  
temperature gradient was measured by two RuO$_{2}$ thermometers connected to 
the "voltage" contacts through the gold wires, and the thermometers were 
supported by long, thin superconducting Nb-Ti wires to minimize heat losses. 
Uncertainties in the absolute magnitudes of $\kappa (T)$ ($\rho (T)$), mainly 
due to the finite contact dimensions on these small crystals, are estimated 
to be around 15$\%$ for $\kappa_{b}$ ($\rho_{b}$) and around 25$\%$ for 
$\kappa_{a}$ ($\rho_{a}$).

The $\kappa (T)$ data for all crystals are shown 
on double-logarithmic axes in Figure 1. The variation of 
$\kappa_{b}(T)$ for the two $\it b$-axis crystals is almost identical over the 
whole temperature range studied, giving us confidence in the reproducibility 
of our data. Below 0.25K, $\kappa_{a}$ and $\kappa_{b}$ both vary approximately 
as $T^{3}$, consistent with phonon heat transport in the boundary-scattering 
limit. Above 0.25K, $\kappa_{a}(T)$ deviates more strongly 
from a $T^{3}$ dependence. The origin of the enhancement of $\kappa_{b}$ over 
$\kappa_{a}$ above $\it T$ = 0.25K is not understood at present, though 
we assume it reflects an additional contribution to $\kappa$ from the CuO 
chains; either QP conductivity on the chains develops swiftly above 
0.25K (note that the CuO chains, being quasi-1D, may be susceptible to charge 
ordering at very low $\it T$), or there exists an additional channel for 
phonon heat propagation\cite{cohn} along the chains that reduces the effects 
of phonon scattering beyond the ballistic regime. Further measurements in 
a magnetic field are envisaged to clarify the origin of this anisotropy. 

In order to look for evidence of a linear $\kappa_{res}$, we have re-plotted 
the low $\it T$ data in Figure 2 as $\kappa_{a,b}/T$ versus $T^{2}$ and 
fitted each data set below 0.25K to the expression $\kappa_{a,b} = AT + 
BT^{3}$\cite{fit}. The coefficients for each fit are $\it A$ = 0.011, - 0.007 
and 0.006 ($\pm$ 0.02) mW/cm.K$^{2}$ and $\it B$ = 6.50, 7.00 and 10.67 ($\pm$ 
0.50) mW/cm.K$^{4}$ for a$\sharp$1, b$\sharp$1 and b$\sharp$2 respectively.

In the boundary-scattering limit, $\kappa_{ph}$ is given by
\begin{equation}
\kappa_{ph} = 1/3 \beta <v_{ph}> l_{0} T^{3}
\end{equation}
where $\beta$ is the phonon specific heat coefficient, $<v_{ph}>$ the average 
acoustic sound velocity and $l_{0} = 2w/\sqrt \pi$ is the maximum phonon mean 
free path. Here, $\it w$ represents a mean width of the rectangular-shaped 
crystal. Taking the dimensions of our crystals and suitable values for 
$\beta$ (= 0.5 $\pm$ 0.1 mJ/mol.K$^{4}$)\cite{nohara} and $<v_{ph}>$ (= 5 
$\pm$ 1 x 10$^{5}$ cm/s)\cite{soundvel}, we obtain estimates for 
$\kappa_{ph}$/$T^{3}$ = 4.1 $\pm$ 1.2, 5.25 $\pm$ 1.5 and 9.55 $\pm$ 2.0
mW/cm.K$^{4}$ for a$\sharp$1, b$\sharp$1 and b$\sharp$2 respectively\cite{sasha}. 
Given the uncertainties in measuring dimensions and contact distances, we 
believe these values compare favourably with the experimental values. 
More importantly, the size of the $T^{3}$ term for the two $\it b$-axis crystals 
scales well with $\it w$ and we conclude that the $T^{3}$ contribution is 
indeed simply the phonon contribution in the ballistic regime.

The most striking result here is the complete absence (to within our experimental 
accuracy) of the residual linear term in the low-$\it T$ $\kappa(T)$ for $\it 
both$ chain and plane current directions. It should be emphasized, of 
course, that a zero linear term in $\kappa_{b}$ also implies a negligible 
$\kappa_{res}$ within the planes, meaning we have effectively confirmed the 
absence of the ÒuniversalÓ QP term in Y124 in all three samples. 
Moreover, for there to be any finite zero-$\it T$ intercept in $\kappa /T$, 
it would require $\kappa_{ph}(T)$ below 0.15K to vary as $\it T^{3+n}$ with $\it 
n$ $>$ 0, which is simply not physical, given that the lattice heat capacity 
is strictly cubic below 1K. Thus, we are confident that
the main result of this Letter, namely the absence of $\kappa_{res}$ in Y124, 
is robust.  For comparison, we also show in the inset to Figure 2, 
$\kappa_{ab}/T$ for optimally doped Bi2212\cite{behnia} $\it measured$ 
$\it with$ $\it the$ $\it same$ $\it experimental$ $\it set$-$\it up$\cite{note}. 
In Bi2212, we can clearly distinguish a finite  $\kappa_{res}/T \approx$ 0.15 
mW/cm.K$^{2}$ (shown by a dotted line), that is $\it an$ $\it order$ $\it of$ 
$\it magnitude$ $\it larger$ than the upper limits for $\kappa_{res}/T$ in Y124.

Despite the overwhelming case for $\it d$-wave pairing in high-$\it T_{c}$ 
cuprates, there is still limited, direct evidence for a $d_{x^{2}-y^{2}}$ 
order parameter in Y124. Thus, before discussing our result in terms of nodal 
QP states, we should first examine the possibility that there is a finite 
gap everywhere on the Fermi surface in Y124. First of all, the orthorhombic distortion 
in Y124, induced by the chains, introduces some $\it d$ + $\it s$ admixture 
in the gap function. As the $\it s$-component is increased from zero, 
the position of the nodal lines are first shifted away from ($\pi, \pi$), but 
as the $\it s$-component becomes comparable with the $\it 
d$-component, a nodeless gap may form. Secondly, magnetic impurities 
are thought to induce a local $\it imaginary$ component that 
could also give rise to a fully gapped state and a suppression of low 
$\it T$ thermal transport\cite{movshovich}. This latter possibility, 
however, is not supported by specific heat data taken on crystals from 
similar batches to those studied here, which show no sign of a low  $\it 
T$ Schottky anomaly arising from such magnetic impurities\cite{nohara}.  
In addition, power law penetration depths have now been observed down to 2K for 
both the $\it a$- and $\it b$-axes in Y124\cite{pana2,broun}, suggesting a 
simple nodal gap picture is equally applicable to Y124. Of course, we cannot 
rule out completely the possibility of a finite gap in Y124, but if it does 
exist, it would have to be vanishingly small. In what follows, 
therefore, we assume that nodal lines are present in Y124 and turn to consider 
what might be happening to the low-energy QP states in their vicinity.

From (1), we see that $\kappa_{res}/T$ is directly proportional to the ratio 
$v_{F}$/$v_{2}$  at the nodal positions, so a negligible 
$\kappa_{res}/T$ may indicate a sharp gap feature at the nodes at very low 
energies, induced either by doping, impurities or structural modifications. 
However, as noted above, $v_{F}/v_{2}$ is expected to $\it increase$ as we move 
into the underdoped regime, and indeed, independent estimates of $v_{F}/v_{2}$ 
from the slope of the low-$\it T$ penetration depth\cite{broun} in Y124 yield 
an estimate for $\kappa_{res}/T$ that is $\it larger$ than those for both 
optimally doped Bi2212 and Y123. Moreover, given that band structure estimates 
for $v_{F}$ are similar for Y123 and Y124\cite{massidda}, even with our upper 
bound estimate for $\kappa_{res}/T$ (= 0.02 mW/cm.K$^{2}$), we obtain a 
physically unrealistic value ($v_{F}/v_{2}$ $\approx$ 2) for the gap slope 
within the nodes in Y124. It appears unlikely, therefore, that the gap 
structure can itself explain a value of $\kappa_{res}/T$ one order of 
magnitude lower than in Y123.

Another important consideration is the size of the impurity band, $\gamma$. 
We recall that in the unitary limit, the universal conductivity regime
develops below $k_{B}T \leq \gamma$. Thus, the observation of  
$\kappa_{res}$ depends not only on the temperature range of the experiment, 
but also on the energy scale of the impurity band, imposed by the scattering 
phase shift.  However, taking values for the $\it b$-axis residual resistivity 
$\rho_{0}$ = 0.5 $\mu \Omega$cm\cite{huss} and plasma frequency $\omega_{p}$ = 
2.5eV\cite{basov}, we obtain a $\it lower$ $\it bound$ estimate of 
$\gamma$ in the unitary limit of 14K, some two 
orders higher than the base temperature of our measurements. (Unfortunately, 
similar analysis for $J//a$ cannot be performed due to the difficulties in 
estimating $\rho_{0}$ from $\rho_{a}(T)$). Even in the opposite (Born) limit, 
where $\gamma$ becomes exponentially small and therefore, may fall below 
our measurement range, the product of the DOS and the lifetime is also energy 
independent above $k_{B}T \geq \gamma$. Hence, $\kappa_{res}/T$ should still 
be constant and finite and the same arguments still apply.

Given these simple yet rather compelling arguments, we are left to consider 
how QP localization might account for the absence of 
$\kappa_{res}/T$ in underdoped Y124. As mentioned above, high magnetic field 
measurements on the underdoped cuprates La$_{2-x}$Sr$_{x}$CuO$_{4}$\cite{ando} 
and Pr$_{2-x}$Ce$_{x}$CuO$_{4}$\cite{fournier} revealed that $\rho_{ab}(T)$, 
though "metallic" at high $\it T$, tends towards localization as $T \rightarrow$ 
0K. The origin of this localization phenomenon is unknown at present. However, 
if the field-induced destruction of superconductivity leads directly to an 
insulating phase, then it is not unreasonable to assume that this transition 
is from a superconductor with already $\it localized$ QP excitations\cite{vish}. 
The crossover from metallic to insulating behavior in the normal state (i.e. 
above $H_{c2}$) as we approach the Mott insulator, suggests an increasing role 
of long-range interactions on the mobility of low-energy quasiparticles. The 
superfluid condensate, suppressed in the vicinity of an impurity, becomes 
ineffective in screening completely the Coulomb repulsion between quasiparticles 
in the bound state\cite{balatsky2}. As we approach the parent insulator, we 
expect such interactions to grow, leading to insulating behavior of the 
quasiparticles above the superconducting condensate and a negligible QP 
contribution to the low-$\it T$ heat transport\cite{PCCO}. Such localization 
in a nominally clean superconductor is an exciting prospect, and measurements 
on Zn-doped or irradiated Y124 are envisaged to investigate this possibility 
further. We note here that most impurity models for $\it d$-wave (cuprate) 
superconductivity fail to take into account the developing role of long-range 
interactions between quasiparticles as the S/I boundary is approached. We 
hope, therefore, that this work stimulates renewed theoretical efforts to 
understand the nature of QP excitations in the CuO$_{2}$ planes, deep inside 
the superconducting state, and in particular in stoichiometric crystals on the 
underdoped side of the phase diagram.

In conclusion, we have measured the low-$\it T$ $\kappa (T)$ of 
stoichiometric, underdoped Y124 and have found that, in marked contrast to 
optimally doped Y123 and Bi2212, the "universal" QP conductivity 
term is absent. We have considered several interpretations of this intriguing 
result, including localization of the quasiparticles themselves, due to enhanced 
long-range interactions as we approach the S/I boundary. Prior to this work, 
the observation of the universal conductivity in Y123 and Bi2212 had been 
widely regarded as solid support for the picture of long-lived 
quasiparticles above the superconducting ground state of high-$\it T_{c}$ 
cuprates. Our surprising result offers important and timely counter 
evidence that the residual conductivity term is non-universal, and we hope 
it encourages further debate and investigation into this critical, and 
still controversial, issue.

We acknowledge enlightening discussions with A.S. Alexandrov, 
J.F. Annett, A.V. Balatsky, D.M. Broun, J.R. Cooper, P.J. Hirschfeld, 
F.V. Kusmartsev, A.P. Mackenzie, A.J. Schofield, L. Taillefer, I. Vekhter 
and V.W. Wittorff. S.N. acknowledges support from the National Science 
Foundation under Grant No.INT-9901436. This work was also partly supported 
by CREST, a Grant in Aid for Scientific Research from the Ministry of 
Education, Science and Culture, Japan and New Energy and Industrial 
Technology Department Organization (NEDO).

\centerline{{\bf Figure Captures}}

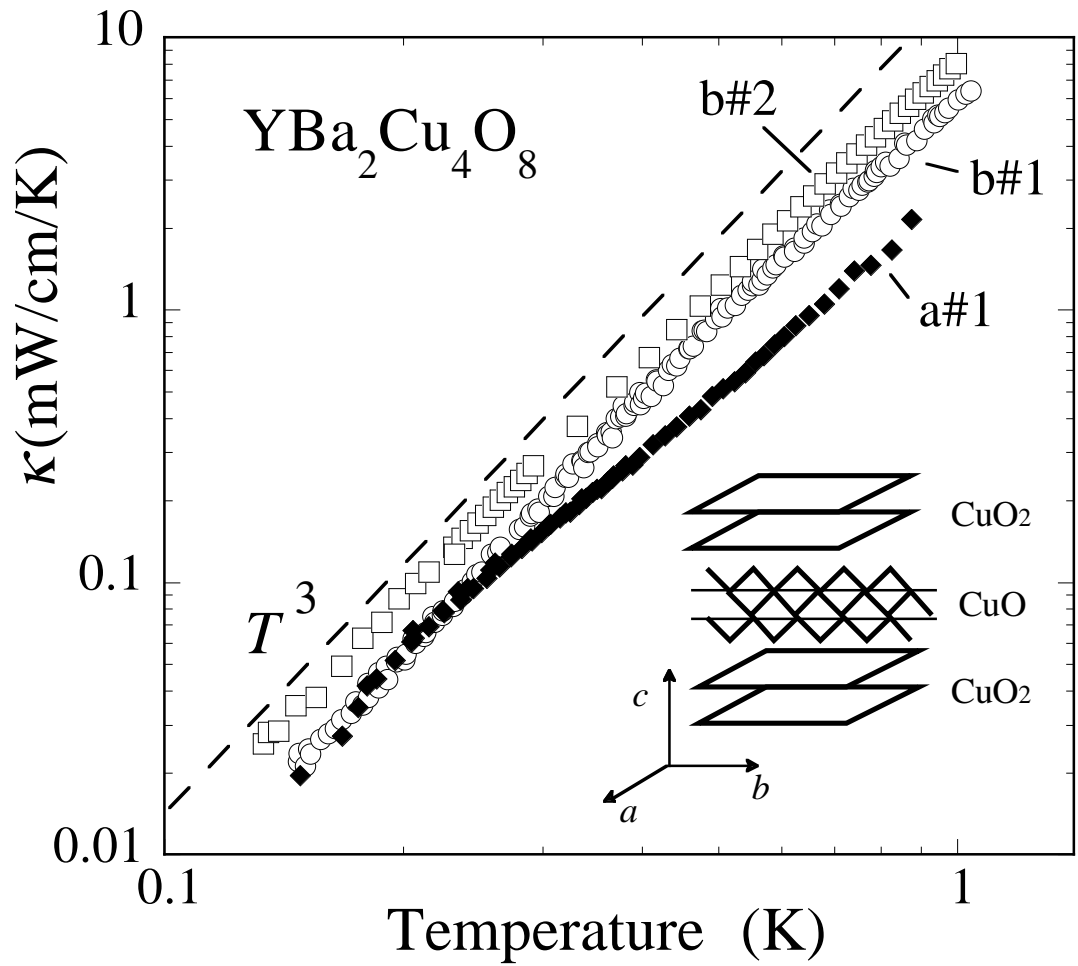
Fig.1. $\kappa_{a}(T)$ and $\kappa_{b}(T)$ of Y124, plotted on 
double-logarithmic axes. The dashed line represents the $T^{3}$ dependence 
expected for phonon heat transport in the ballistic regime. The inset 
shows a simple schematic of the crystal structure of Y124.

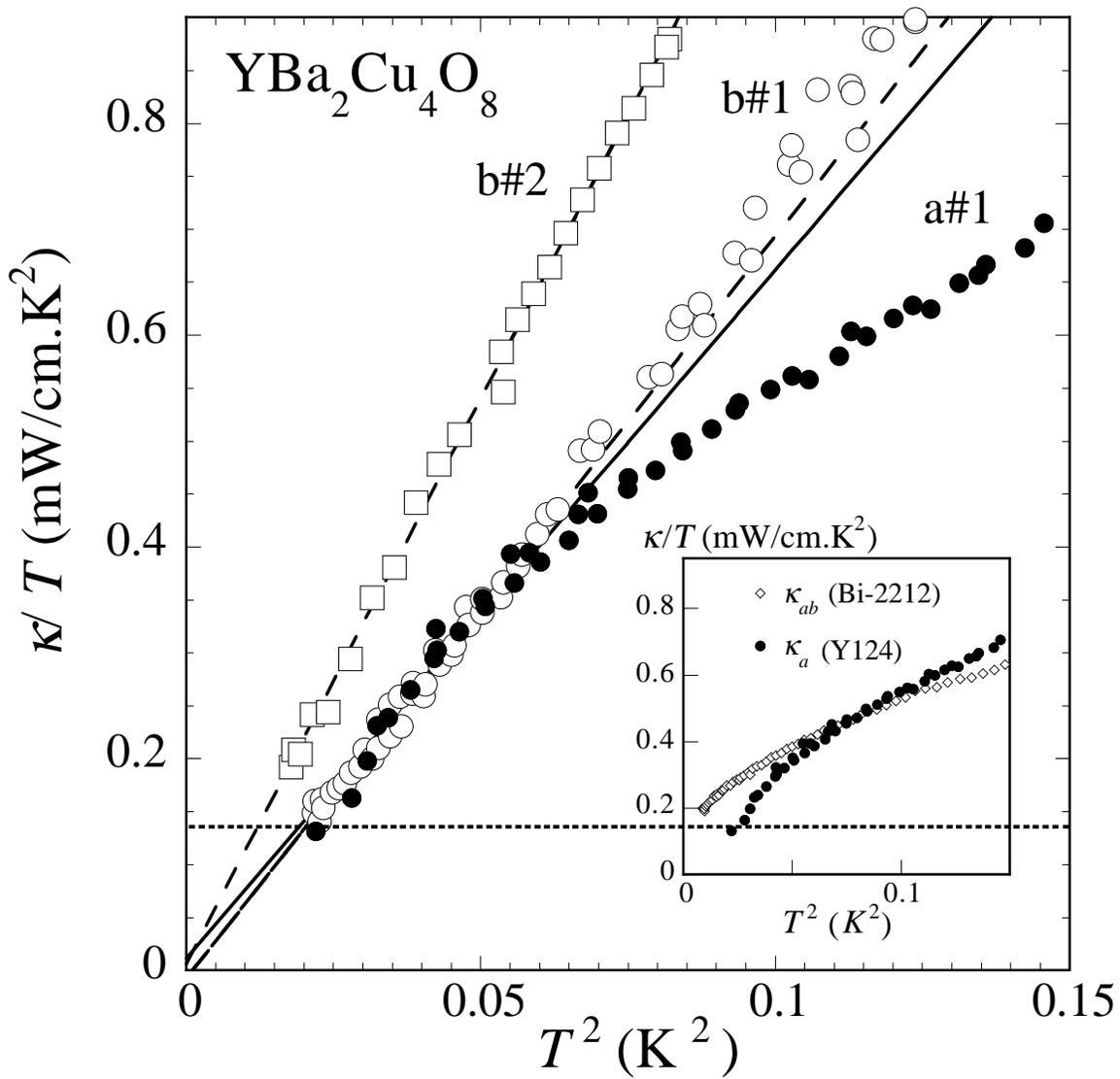
Fig.2. $\kappa/T$ versus $T^{2}$ below 0.4K for a$\sharp$1 (closed circles), 
b$\sharp$1 (open circles) and b$\sharp$2 (open squares). Fits to the expression 
$\kappa = AT + BT^{3}$ below 0.25K are indicated by dashed lines for 
b$\sharp$1 and b$\sharp$2 and by a solid line for a$\sharp$1. The dotted line 
represents the universal conductivity limit for Bi2212. Inset: Comparison of 
$\kappa_{a}/T$ for Y124 (closed circles) and $\kappa_{ab}/T$ of 
Bi2212\cite{behnia} (open diamonds), measured in the same apparatus.


\begin{thebibliography}{99}
\bibitem{note}
Similar values are found for Y123\cite{chiao}.
\end{thebibliography}
\end{document}